\documentclass[conference,10pt]{IEEEtran}
\IEEEoverridecommandlockouts
\usepackage{cite}
\usepackage{amsmath,amssymb,amsfonts}
\usepackage{algorithmic}
\usepackage{graphicx}
\usepackage{textcomp}
\usepackage{xcolor}
\usepackage[capitalise, noabbrev]{cleveref}
\usepackage{amsmath}
\usepackage{subcaption}
\usepackage{multirow}
\usepackage{balance}
\usepackage{ctable}
\usepackage{threeparttable}
\def\BibTeX{{\rm B\kern-.05em{\sc i\kern-.025em b}\kern-.08em
    T\kern-.1667em\lower.7ex\hbox{E}\kern-.125emX}}
\begin{document}

% Human Activity Recognition from Irregularly Sampled Data Using Liquid Time-Constant Neural Networks: Lessons Learned
\title{Assessing the Impact of Sampling Irregularity in Time Series Data: Human Activity Recognition As A Case Study}

% \author{\IEEEauthorblockN{1\textsuperscript{st} Given Name Surname}
% \IEEEauthorblockA{\textit{dept. name of organization (of Aff.)} \\
% \textit{name of organization (of Aff.)}\\
% City, Country \\
% email address or ORCID}
% \and
% \IEEEauthorblockN{2\textsuperscript{nd} Given Name Surname}
% \IEEEauthorblockA{\textit{dept. name of organization (of Aff.)} \\
% \textit{name of organization (of Aff.)}\\
% City, Country \\
% email address or ORCID}
% \and
% \IEEEauthorblockN{3\textsuperscript{rd} Given Name Surname}
% \IEEEauthorblockA{\textit{dept. name of organization (of Aff.)} \\
% \textit{name of organization (of Aff.)}\\
% City, Country \\
% email address or ORCID}
% \and
% \IEEEauthorblockN{4\textsuperscript{th} Given Name Surname}
% \IEEEauthorblockA{\textit{dept. name of organization (of Aff.)} \\
% \textit{name of organization (of Aff.)}\\
% City, Country \\
% email address or ORCID}
% \and
% \IEEEauthorblockN{5\textsuperscript{th} Given Name Surname}
% \IEEEauthorblockA{\textit{dept. name of organization (of Aff.)} \\
% \textit{name of organization (of Aff.)}\\
% City, Country \\
% email address or ORCID}
% \and
% \IEEEauthorblockN{6\textsuperscript{th} Given Name Surname}
% \IEEEauthorblockA{\textit{dept. name of organization (of Aff.)} \\
% \textit{name of organization (of Aff.)}\\
% City, Country \\
% email address or ORCID}
% }

\author{\IEEEauthorblockN{
        Mengxi Liu\IEEEauthorrefmark{1},
        Daniel Geißler\IEEEauthorrefmark{1},
        Sizhen Bian \IEEEauthorrefmark{1},
        Bo Zhou\IEEEauthorrefmark{1}\IEEEauthorrefmark{2},and
        Paul Lukowicz\IEEEauthorrefmark{1}\IEEEauthorrefmark{2}
        }
    \IEEEauthorblockA{\IEEEauthorrefmark{1}German Research Center for Artificial Intelligence (DFKI), Kaiserslautern, Germany}
    \IEEEauthorblockA{\IEEEauthorrefmark{2}Department of Computer Science, RPTU Kaiserslautern-Landau, Kaiserslautern, Germany}
    \IEEEauthorblockA{Email: firstname.lastname@dfki.de} 
}

% \IEEEpeerreviewmaketitle
% \setcounter{page}{1}
\maketitle

\begin{abstract}
Human activity recognition (HAR) ideally relies on data from wearable or environment-instrumented sensors sampled at regular intervals, enabling standard neural network models optimized for consistent time-series data as input. However, real-world sensor data often exhibits irregular sampling due to, for example, hardware constraints, power-saving measures, or communication delays, posing challenges for deployed static HAR models. This study assesses the impact of sampling irregularities on HAR by simulating irregular data through two methods: introducing slight inconsistencies in sampling intervals (timestamp variations) to mimic sensor jitter, and randomly removing data points (random dropout) to simulate missing values due to packet loss or sensor failure.
We evaluate both discrete-time neural networks and continuous-time neural networks, which are designed to handle continuous-time data, on three public datasets. 
We demonstrate that timestamp variations do not significantly affect the performance of discrete-time neural networks, and the continuous-time neural network is also ineffective in addressing the challenges posed by irregular sampling, possibly due to limitations in modeling complex temporal patterns with missing data. Our findings underscore the necessity for new models or approaches that can robustly handle sampling irregularity in time-series data, like the reading in human activity recognition, paving the way for future research in this domain.

\end{abstract}


% \vspace{-1em}
\section{Introduction}

The integration of Artificial Neural Networks (ANNs) into mobile systems has gained significant traction as a research direction, aligning with the contemporary trend of migrating artificial intelligence (AI) computations closer to the data sources \cite{sipola2022artificial}.
% This trend is driven by the need for efficient processing and reduced latency in various applications. 
However, deploying ANNs on edge devices poses challenges that extend beyond computational complexity \cite{singh2023survey}.
Specifically, the inherent variability in sampling rates across diverse edge devices and sensors, operating under disparate conditions, necessitates either extensive model re-tuning with supplementary data or the development of more sophisticated models designed to accommodate the adaptability \cite{hasegawa2020smartphone}. 
In reality, sampling rates of sensors additionally vary unpredictably due to unexpected factors such as hardware scheduling and interrupts \cite{hamouda2020variable}.
Irregular sampling rates in mobile systems can arise from a multitude of factors, including sensor-related issues such as noise, errors, saturation, and limited range \cite{coviello2020study}; network-related problems like communication link failures, packet loss or corruption, and congestion \cite{chinaev2021online,vasconcelos2018data}; system-related constraints like power management, and operating system scheduling \cite{he2006sensor, zhang2013distributed}; 
% environmental factors including interference, physical obstacles, and weather conditions \cite{rahimi2005adaptive}; 
% human-related aspects like user behavior, mobility, device handling, and neglect of maintenance and calibration \cite{khan2016optimising}; 
and design-related limitations such as insufficient margins, lack of robustness in algorithm design, and inadequate testing \cite{wang2010networked}. 
These diverse factors can interact and compound each other, making it challenging to maintain a consistent sampling rate and requiring careful consideration and mitigation strategies to ensure reliable data collection and system performance. 
However, mainstream ANNs powering such intelligent mobile systems typically assume a uniform time granularity of input samples \cite{ananthanarayanan2017real, liu2024ikan, bian2022state}, an assumption which is often over-optimistic when developing on GPU-based infrastructure with pre-processed datasets that do not reflect the reality of mobile systems. 

% Sensor-based human activity recognition (HAR) provides a practical example of leveraging embedded ANNs. 
Nonetheless, recent advancements in sensor-based HAR have yielded significant improvements across various fields, including gesture recognition and health monitoring \cite{bian2021capacitive, phukan2022convolutional, chen2021deep}. 
For wearable and mobile activity recognition applications, energy efficiency and battery life are paramount to ensure a seamless user experience.
Notably, also the energy demand for data transmission varies with complex tasks requiring higher sampling rates, whereas simpler activities can be accurately identified at lower rates \cite{liu2018impact}. 
Consequently, an adaptive sampling rate strategy is essential for achieving energy efficiency in edge device applications \cite{qi2013adasense}.   % adaptive sampling rate algorithms?
However, modeling time series sensor data with non-uniform intervals remains a formidable challenge for discrete ANNs such as Recurrent Neural Networks (RNNs) \cite{rubanova2019latent}.

% Several approaches have been proposed to handle irregular time-series data, including imputation methods to estimate missing values and the development of continuous-time neural network architectures \cite{che2018recurrent, rubanova2019latent}. 
As a promising solution, Liquid Time-Constant Neural Networks (LTNNs) obtain an architecture designed to process continuous-time signals and theoretically accommodate irregular sampling \cite{hasani2021liquid}.
LTNNs adjust their internal time constants based on the input data, allowing them to model temporal dynamics more flexibly than their discrete-time counterparts.
Despite these advancements, there remains a gap in understanding how well such models perform under different types of sampling irregularities commonly encountered in HAR tasks. 
Moreover, there is a need to systematically evaluate the impact of such irregularities on the performance of both traditional discrete-time neural networks and continuous-time models like LTNNs. Within this work, we evaluate the impact of irregular sampling on HAR by simulating irregularities in otherwise regular datasets through two methods: (1) Timestamp Variations, which introduce minor interval inconsistencies, and (2) Random Dropout, which removes data points to emulate missing values. Using three publicly available HAR datasets, we investigate the usefulness of both, discrete-time neural networks and LTNNs. 
% Our results show that while discrete-time networks remain robust to slight sampling variations, LTNNs struggle with the more severe irregularities introduced by random dropout. These findings underscore the need for new or adapted models to handle irregular time-series data. Future work could explore advanced continuous-time architectures, better data imputation techniques, or hybrid models to improve HAR reliability in real-world settings.

\section{Related Work}

% Human Activity Recognition (HAR) using wearable sensors has been extensively studied over the past decade, with significant advancements in sensor technology and machine learning algorithms. Traditional HAR systems rely on data sampled at regular intervals, enabling the use of standard neural network architectures optimized for consistent time-series data processing \cite{lara2013survey, ordonez2016deep}. However, the assumption of regular sampling does not hold in many real-world applications due to various factors such as hardware limitations, energy constraints, and communication delays \cite{kwapisz2011activity, chen2015deep}. This section reviews existing approaches to handling irregularly sampled time-series data in HAR and related domains, highlighting their advantages and limitations concerning our study.

% Handling Missing Data and Irregular Sampling in Time Series
Interpolation is a common approach for dealing with irregular sampling at the data level, where missing values are estimated to reconstruct a regularly sampled time series. Methods such as linear interpolation, spline interpolation, and statistical imputation have been widely used \cite{schafer2002missing}. 
While interpolation can be effective for small, random amounts of missing data, it may introduce biases or distort temporal dynamics when dealing with irregularities obtaining extensive durations \cite{nelson1997modeling} and inefficiency during training \cite{geissler2024power}.
% Moreover, imputation methods often assume that missing data points are random, which may not be the case in practical scenarios.
At the model level, Recurrent Neural Networks (RNNs) have been adapted to handle missing data by incorporating masking and time interval information into their architectures. 
Che et al. \cite{che2018recurrent} proposed the GRU-D model, which modifies the Gated Recurrent Unit (GRU) to account for missing values and irregular time intervals by decaying the hidden states. 
This approach allows the model to learn temporal patterns despite irregularities in the data. However, GRU-D and similar models primarily focus on healthcare data and may not generalize well to HAR tasks, where sensor data can be more complex and exhibit different patterns of irregularity.
Continuous-time models, such as Neural Ordinary Differential Equations (Neural ODEs), have been introduced to handle irregularly sampled data by modeling the continuous dynamics underlying the observations \cite{chen2018neural, rubanova2019latent}. 
These models define the hidden state dynamics as a continuous function parameterized by neural networks, naturally allowing them to process data with irregular time stamps. 
Rubanova et al. \cite{rubanova2019latent} extended Neural ODEs to latent variable models for time series with irregular sampling.
Liquid Time-Constant Neural Networks (LTNNs) have been proposed as an alternative continuous-time model that adapts the time constants of neurons based on input data, aiming to capture temporal dynamics more effectively \cite{hasani2021liquid}. 
LTNNs adjust their internal dynamics in response to incoming signals, potentially making them suitable for processing irregularly sampled time series. 
Hasani et al. \cite{hasani2021liquid} demonstrated LTNNs' capability in modeling complex temporal patterns and suggested their applicability to various time-series tasks. Despite their theoretical advantages, their performance in the presence of significant amounts of missing data or highly irregular sampling intervals has not been thoroughly evaluated. Therefore, we proposed this work to comprehensively assess the impact of sampling irregularity on time series data using both discrete and continuous time ANNs, and with the HAR applications as the target case study.

\section{Methodology}

\begin{figure*}[ht]
\footnotesize
\centering
     \begin{subfigure}[b]{0.49\textwidth}
         \centering
         \includegraphics[width=0.8\textwidth]{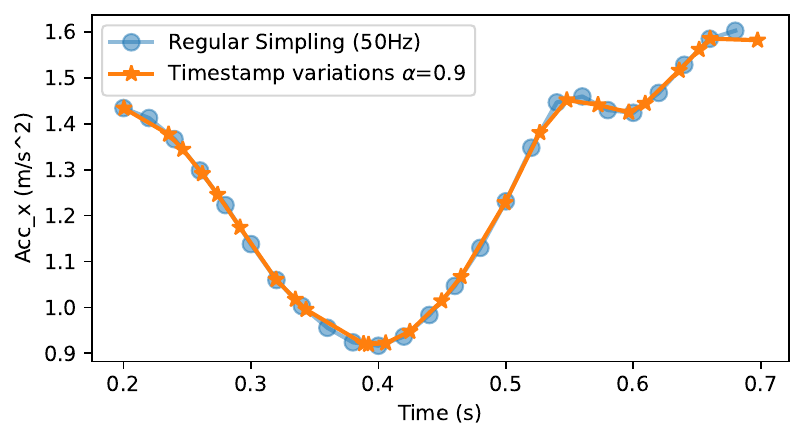}
         \vspace{-1em}
         \caption{Samples with timestamp}
         \label{fig:irregularity_wt}
     \end{subfigure}
     \hfill
     \begin{subfigure}[b]{0.49\textwidth}
         \centering
         \includegraphics[width=0.8\textwidth]{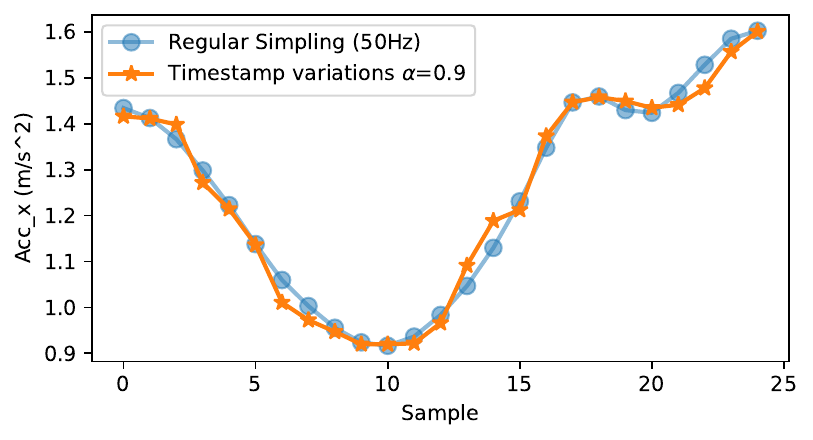}
         \vspace{-1em}
         \caption{Samples without timestamp}
         \label{fig:irregularity_wo}
     \end{subfigure}
    \caption{Signal example of the results from timestamp variation.}
    \label{fig:time_variation}
    \vspace{-1em}
\end{figure*}

\begin{figure*}[ht]
\footnotesize
\centering
     \begin{subfigure}[b]{0.49\textwidth}
         \centering
         \includegraphics[width=0.8\textwidth]{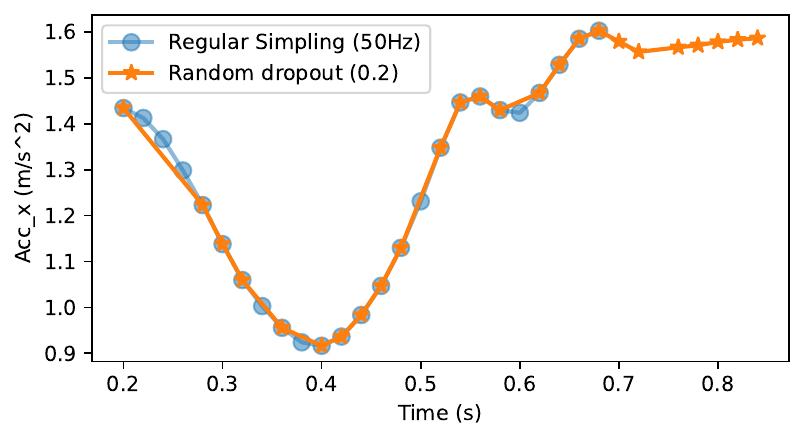}
         \vspace{-1em}
         \caption{Samples with timestamp}
         \label{fig:irregularity_wt_dropout}
     \end{subfigure}
     \hfill
     \begin{subfigure}[b]{0.49\textwidth}
         \centering
         \includegraphics[width=0.8\textwidth]{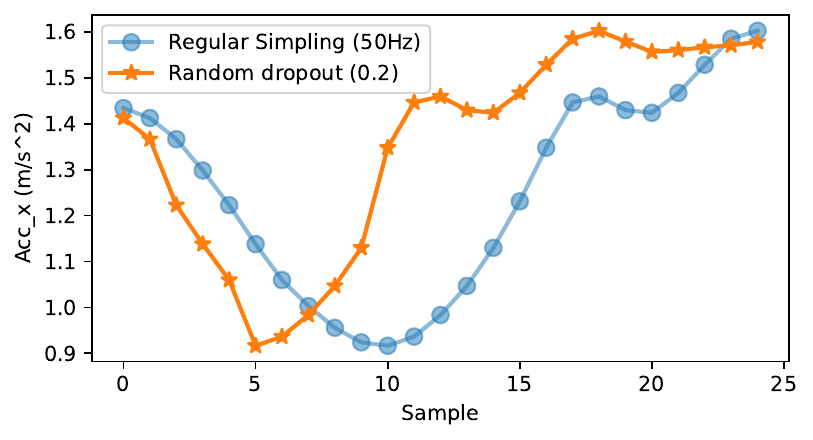}
         \vspace{-1em}
         \caption{Samples without timestamp}
         \label{fig:irregularity_wo_dropout}
     \end{subfigure}
     
    \caption{Signal example of the results from random dropout}
    \label{fig:dropout}
    \vspace{-1em}
\end{figure*}

% In this work, we evaluate the impact of sampling irregularity in HAR tasks, however, 
Most of the existing research about HAR benchmarking was conducted on public datasets with the assumption of sensor data sampling under the constant time interval, with datasets like PAMAP2 \cite{reiss2012pamap2}, MHEALTH \cite{banos2014mhealth} and MotionSense \cite{malekzadeh2018motionsense}.
However, sampling irregularity are often ignored, wherefore we propose two methods to simulate the potential sampling irregularity occurring on realistic occasions:

\begin{itemize}
    \item \textbf{Timestamp variations}: Introducing slight inconsistencies in sampling intervals to mimic sensor jitter.
    \item \textbf{Random dropout}: Randomly removing data points to simulate missing values due to packet loss or sensor failure.
\end{itemize}

\textbf{Timestamp Variation:} Given a regularly sampled time series $D = \{ (t_i, x_i, y_i) \mid i = 1, 2, \ldots, N \}$,  where $ t_i $ is the timestamp of sample $i$,
${x_i \in \mathbb{R}^d}$ represents the feature vector corresponding to sample $ i$ (with  $d$  features), $y_i$ is the label corresponding to sample $i$.
The original timestamps of the time series with a fixed sampling interval $\Delta t$, i.e.,
\begin{equation}
    t_i = t_1 + (i-1) \Delta t
\end{equation}
The goal of the timestamp variations method is to generate a set of irregular timestamps $t'_i$, where:
\begin{equation}
    t'_i = t_i + \delta_i, \quad i = 1, 2, \ldots, N
\end{equation}
$\delta_i$ represents a random offset for each timestamp, here sampled from a uniform distribution $ \delta_i \sim \mathcal{U}(-\epsilon, \epsilon) $, where $ \epsilon $ is a small value that controls the magnitude of the irregularity.
We defined the $ \epsilon $ as the timestamp variation rate.
Thus, the irregular timestamps are:
\begin{equation}
    t'_i = t_1 + (i-1) \Delta t + \delta_i
\end{equation}
After the generation of the new irregular timestamps, the next step is to find the corresponding values at the new irregular timestamps $ t'_i $:
to obtain $ x'_i $, the new value of the time series at the irregular time \( t'_i \), the linear interpolation was implemented in this work:
\begin{equation}
    x'_i = X(t'_i)
\end{equation}
Where  $X(t)$  is an interpolation function based on the original time series values $ (t_i, x_i)$. 
For linear interpolation, $ X(t) $ is defined piecewise between each consecutive pair of points $(t_i, x_i)$ and $(t_{i+1}, x_{i+1})$:
\begin{equation}
    X(t) = x_i + \frac{(x_{i+1} - x_i)}{(t_{i+1} - t_i)} (t - t_i), \quad t_i \leq t \leq t_{i+1}
\end{equation}
The new values $ x'_i $ can be obtained for the irregular timestamps $ t'_i $ by the following interpolation function:
\begin{equation}
    x'_i = X(t'_i)
\end{equation}
Thus, the resulting irregularly sampled time series is represented by the pairs $ (t'_i, x'_i) $, where $ i = 1, 2, \ldots, N $.
This process converts a regular time series into an irregular one, preserving the temporal order while introducing irregular spacing between samples. The label $y_i$ remains the same.

\textbf{Random Dropout:}
We further implemented the random dropout method to simulate the packet loss or sensor failure occasions with unavailable data and reduced number of samples.
We defined the random dropout rate $\alpha$, the number of the samples to drop is $N_{\text{drop}} = \lfloor \alpha \times N \rfloor$, where $ \lfloor \cdot \rfloor $ denotes the floor function to ensure an integer value.
Next, randomly selecting $ N_{\text{drop}} $ indices from the set  $\{ 1, 2, \ldots, N \}$  without replacement: $I_{\text{drop}} \subset \{ 1, 2, \ldots, N \}, \quad |I_{\text{drop}}| = N_{\text{drop}}$.
Then, removing the samples with the selected indices from the time series dataset:
\begin{equation}
    D' = D \setminus \{ (t_i, x_i, y_i) \mid i \in I_{\text{drop}} \}
\end{equation}
where $D'$ is the resulting dataset after dropping the selected elements.

\cref{fig:time_variation} and \cref{fig:dropout} present the signal comparison between the original regular sampling data and irregular signal resulting produced by the timestamp variation and random dropout methods. 
Two kinds of X-axis were used to visualize the signals, such as the actual time in seconds and sample indices, highlighting that the timestamp information plays an important role in signal representation, which is only considered in continual time ANNs, while the standard ANNs ignore it. 
As shown in \cref{fig:irregularity_wt} and \cref{fig:irregularity_wt_dropout}, the trend of the irregular sampling signal caused by timestamp variation and random dropout almost retains the same as the regular sampling signal when the timestamp information is added into the signal visualization.
However, there is an obvious variation between the regular and irregular sampling signal when time information is not available as \cref{fig:irregularity_wo} and \cref{fig:irregularity_wo_dropout}.
In addition, it can be observed that the random drop method can cause more intensive signal variation than the timestamp variation. 

\section{Evaluation}
\begin{table*}
\renewcommand{\arraystretch}{0.8}
\centering
\caption{Average Performance Loss Results with Different Timestamp Variation (Sampling rate of the training dataset is 50 Hz)}
\begin{tabular}{lccccccccccc}
\toprule
\multirow{2}{*}{Dataset} & \multirow{2}{*}{Model} & \multirow{2}{*}{Macro F1 Score}& \multicolumn{9}{c}{Performance Loss with 
 Different Timestamp Variation (\%)}\\
  &  &  & 0.1 & 0.2 & 0.3 & 0.4 & 0.5 & 0.6 & 0.7 & 0.8 & 0.9 \\
\midrule
\multirow{4}{*}{PAMAP2} 
& ConvDense & 0.622 & 0.34 & -0.53 & 0.84 & 0.05 & 1.33 & 1.38 & 0.88 & 0.77 & 1.79 \\
& DeepConvLSTM & 0.582 & 0.35 & 0.71  & 1.07 & 1.11 & 1.07 & 1.42 & 3.77 & 1.57 & 2.80 \\
& TinyHAR & 0.647 & -2.28 & -2.26 & -2.28 & -1.97 & -3.03 & -2.66 & -2.46 & -2.37 & -2.70 \\
& CFC\_Solver & 0.573 & 0.15 & 0.29  & 0.48 & 0.45 & 0.78 & 0.96 & 1.37 & 1.70 & 1.97 \\
\midrule
\multirow{4}{*}{MHEALTH} 
& ConvDense & 0.635 & -0.27 & -0.60 & -0.35 & -0.21 & -0.03 & 0.06 & -0.03 & 0.69 & 0.24 \\
& DeepConvLSTM & 0.600 & -0.26 & -0.52 & -0.85 & -0.40 & -0.68 & -0.48 & -0.48 & 0.31 & 0.31 \\
& TinyHAR & 0.651 & 0.03  & 0.27  & 0.35  & 0.52  & 0.62  & 0.68  & 0.77  & 0.85  & 0.52 \\
& CFC\_Solver & 0.605 & -0.08 & -0.18 & -0.20 & -0.07 & -0.11 & 0.03  & 0.01  & 0.16  & 0.07 \\
\midrule
\multirow{4}{*}{MOTIONSENSE} 
& ConvDense & 0.849 & 0.11 & 0.24 & 0.34 & 0.37 & 0.55 & 0.56 & 0.86 & 0.72 & 0.62 \\
& DeepConvLSTM & 0.870 & 0.12 & 0.24 & 0.38 & 0.48 & 0.56 & 0.49 & 1.04 & 1.06 & 0.82 \\
& TinyHAR & 0.871 & -0.10 & 0.01 & -0.09 & 0.16 & 0.14 & 0.24 & 0.43 & 0.44 & 0.72 \\
& CFC\_Solver & 0.819 & -0.04 & -0.05 & 0.02 & -0.01 & -0.00 & 0.10 & 0.12 & 0.23 & 0.28 \\

\bottomrule
\label{tab:result_summary}
\vspace{-2em}
\end{tabular}
\end{table*}

\begin{figure*}[ht]
\footnotesize
\centering
 \includegraphics[width=0.8\textwidth]{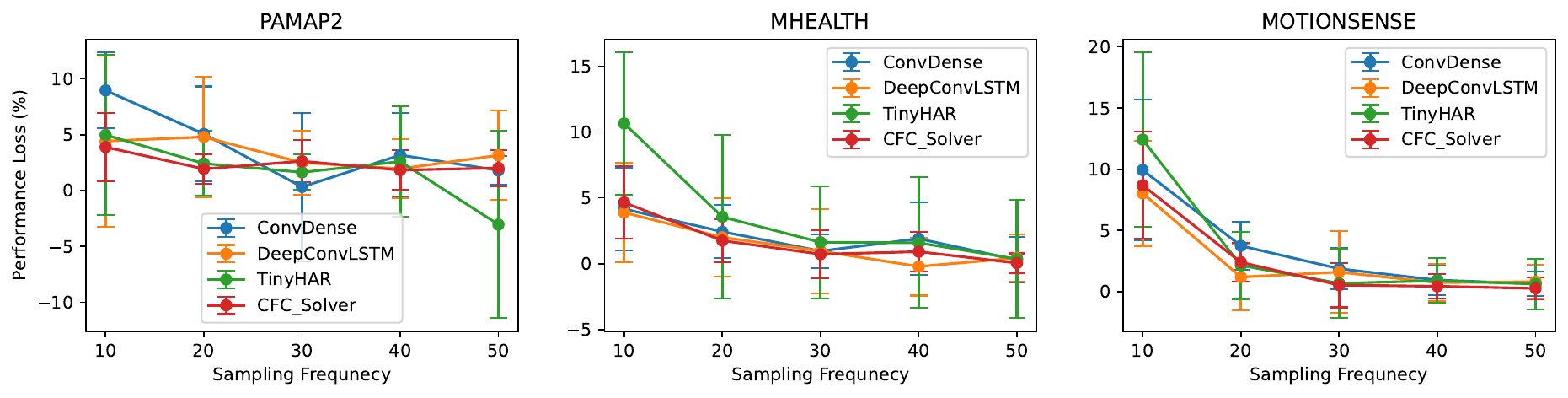}
 \caption{Performance loss caused by timestamp variation of 0.9 when training the model with different sampling rate datasets}
 \label{fig:performance_loss_sampling_rate}
 % \vspace{-1em}
\end{figure*}

\begin{figure*}[ht]
\footnotesize
\centering
 \includegraphics[width=0.8\textwidth]{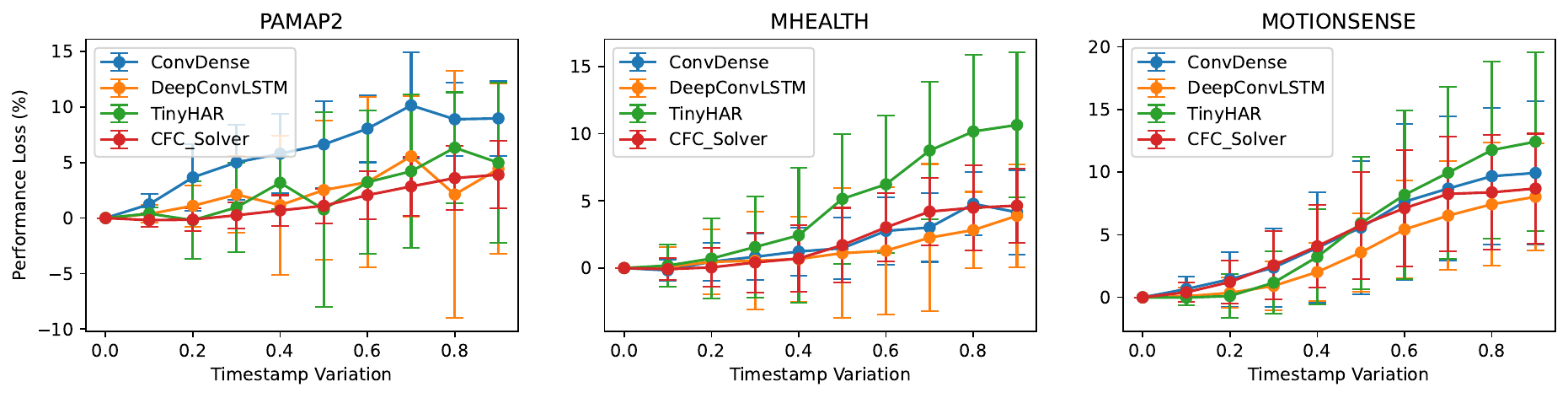}
 \caption{Performance loss with different timestamp variation (The sampling rate of the training dataset is 10 Hz)}
 \label{fig:performance_loss_timevariation}
 \vspace{-1em}
\end{figure*}

\begin{figure*}[ht]
\footnotesize
\centering
 \includegraphics[width=0.8\textwidth]{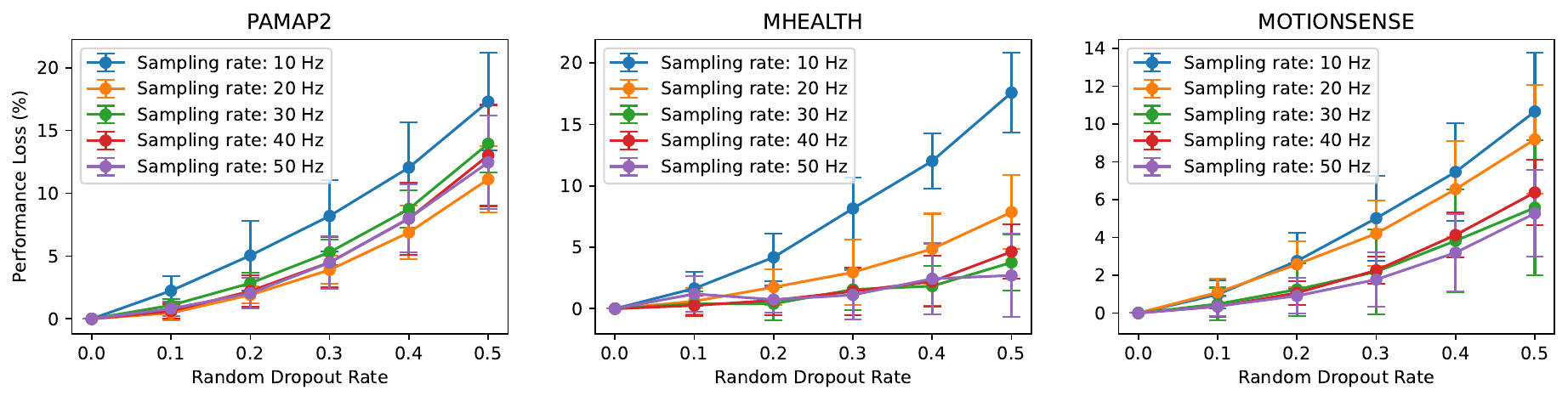}
 \caption{Performance loss of the CFC\_Solver model caused by sample random dropout (Tested with different sampling rates)}
 \label{fig:performance_loss_dropout}
 \vspace{-1em}
\end{figure*}

\textbf{Datasets:}
We used three public datasets to evaluate the impact of sampling irregularity on Human Activity Recognition (HAR) tasks: PAMAP2 \cite{reiss2012pamap2}, MHEALTH \cite{banos2014mhealth}, and MotionSense \cite{malekzadeh2018motionsense}.

\textit{PAMAP2:} In this study, we used only the motion sensor data from the PAMAP2 dataset, creating a customized dataset with 18 channels (three 3-axis accelerometers and three 3-axis gyroscopes) from seven subjects. The dataset contains 12 activity classes, such as walking, running, cycling, and ironing, along with a null class. The original sampling rate is 100 Hz.

\textit{MHEALTH:} For the MHEALTH dataset, all signal channels except the electrocardiogram were used, spanning data from 10 subjects. It includes 13 daily human activities (with a null class), such as climbing stairs and jumping. The original sampling rate is 50 Hz. Since timestamps are not provided in this dataset, we generated timestamps, as they are necessary for training continual-time artificial neural networks (ANNs).

\textit{MotionSense:} The MotionSense dataset includes 12 sensor channels with a sampling rate of 50 Hz, providing rotation, Euler angle, and acceleration information. Data from the first 10 subjects was utilized, covering six activity classes. Similar to MHEALTH, we added manually generated timestamps to this dataset.

\textbf{Neural Networks:}
We evaluated three popular discrete-time ANNs for HAR tasks, namely DeepConvLSTM \cite{ordonez2016deep}, TinyHAR, and ConvDense, and one continual-time ANN, LTC-NN \cite{hasani2021liquid}, to assess the sampling irregularity impact.

\textit{DeepConvLSTM:} This model is a general deep framework for activity recognition, combining convolutional and LSTM recurrent units to explicitly model the temporal dynamics of feature activations. Our implementation includes four one-dimensional convolutional layers, one LSTM layer, and one linear layer.

\textit{TinyHAR:} TinyHAR is an efficient deep learning model for sensor-based HAR tasks. It integrates convolutional and transformer layers to extract spatial information from multi-sensor channels and LSTM layers to enhance temporal information. We used this model directly to recognize activities from the datasets, leveraging its proven efficiency and performance.

\textit{ConvDense:} This model omits recurrent neural network (RNN) layers, focusing solely on spatial information. It consists of three one-dimensional convolutional layers followed by two linear layers.

\textit{LTC-NN:} LTC-NN is a continual-time recurrent neural network model based on linear first-order dynamical systems modulated through nonlinear gates. Outputs are computed using numerical differential equation solvers, with time constants adapting to hidden states. We employed an efficient closed-form solution (referred to as \textbf{CFC\_Solver}) instead of numerical solvers, as proposed in \cite{hasani2022closed}. This model has demonstrated superior performance in addressing sampling irregularity compared to other models such as RNN-Impute \cite{rubanova2019latent} and GRU-D \cite{che2018recurrent}. 
However, previous studies evaluated this model on one small HAR dataset only, motivating its selection for our study.

A sliding window with the length of 2 seconds and step size of 1 second was generated as the input of the neural networks.

\textbf{Validation Method:}
We used leave-one-person-out cross-validation to evaluate model performance. In each iteration, data from one individual served as the test dataset, while data from the remaining individuals formed the training set. A subset of the training data was randomly selected for validation. The macro F1 score was chosen as the evaluation metric. Performance loss \((P_{\text{loss}})\) was calculated as:
\[
P_{\text{loss}} = \frac{P_{\text{regular}} - P_{\text{irregular}}}{P_{\text{regular}}}
\]
where \(P_{\text{regular}}\) is the macro F1 score on regularly sampled data, and \(P_{\text{irregular}}\) is the score on irregularly sampled data.

\textbf{Test Strategy:}
Given that training data collection can occur under controlled conditions (e.g., in a laboratory) with higher data quality than real-world scenarios, we trained the models on regularly sampled datasets and tested their performance on simulated irregular datasets generated using our proposed methods.
Additionally, we investigated the effect of training dataset sampling rates on HAR performance under irregular sampling during testing. Public datasets were down-sampled from their original frequencies to 10 Hz, with steps of 10 Hz.

\textbf{Results:} \cref{tab:result_summary} presents a summary of the average performance loss incurred by various HAR models when subjected to increasing levels of timestamp variation in the test data. All models were trained on datasets sampled at 50 Hz regularly, and the test datasets were artificially altered to introduce sampling irregularities by varying timestamps. Across all conditions, the impact of timestamp variation on HAR performance is generally modest, with performance losses remaining below 3\%. As the level of timestamp variation increases, the severity of performance degradation tends to rise, indicating a direct relationship between timestamp irregularity and reduced HAR accuracy. Nevertheless, some unexpected improvements are also observed; for example, certain moderate timestamp variations appear to enhance the discriminatory patterns for specific activities. This result is particularly obvious in the TinyHAR model when evaluated on the PAMAP2 dataset, where slight timestamp alterations lead to marginal performance gains rather than losses.
In terms of model comparisons, the CFC\_Solver typically shows lower performance degradation than the other models under high timestamp variation, except in the case of the PAMAP2 dataset. However, despite its robustness to timestamp irregularities, the CFC\_Solver often achieves a lower overall macro F1 score than its counterparts. 
This suggests that while CFC\_Solver may handle irregular temporal input more gracefully, it may not always reach the same peak performance as other models under ideal conditions (regular sampling inputs).

Since the sampling frequency of the training dataset can influence the model’s robustness to timestamp variation, we conducted experiments using datasets sampled at several rates, decreasing from 50 Hz to 10 Hz in steps of 10 Hz. \Cref{fig:performance_loss_sampling_rate} shows the performance loss results at a fixed timestamp variation rate of 0.9. The results indicate that models trained on higher sampling rates demonstrate better robustness to timestamp variation during inference.
Focusing specifically on models trained at 10 Hz, we observe an interesting comparison between time-discrete DeepConvLSTM and the time-continuous CFC\_Solver model. In both the MHEALTH and MotionSense datasets, DeepConvLSTM exhibits lower performance loss than CFC\_Solver, despite being a time-discrete model that does not explicitly leverage timestamp information. This suggests that, under certain conditions, a time-discrete model can outperform a time-continuous model in terms of robustness when faced with significant timestamp variation.

\cref{fig:performance_loss_timevariation} shows the performance loss caused by different timestamp variation with sampling rate of the training dataset set to 10 Hz.
It can be observed that the average performance loss has more significant increase than the model trained on 50 Hz data when the timestamp variation rate is increased. 
Among all experiements, the biggest performance loss often occurred with the largest timestamp variation by the TinyHAR model, showing its low robustness to the irregular dataset, although it achieved the highest macro F1 score as shown in \cref{tab:result_summary}.

As the timestamp variation method still keeps and only moves the sampling data timestamps around, the random dropout method was implemented to simulate the complete sample loss. 
We fixed the window size in 2 seconds, whereas the feature number within the window can be varied along to the random dropout rate.
Noteworthy, different from the discrete time ANNs, the CFC\_Solver can process the input instance with flexible length of the window size. 
\cref{fig:performance_loss_dropout} presents the results of the experiment, indicating that random dropout in test data adversely affects model performance across different wearable-sensor datasets. 
Although all models suffer some degree of performance loss when sensor data is randomly removed, the impact is not uniform. 
Higher data sampling rates during training tend to provide a buffer against the negative effects of dropout, helping to maintain more stable performance even as more data points are lost. 
Therefore, optimizing sampling rates and implementing strategies to handle missing data effectively can enhance model resilience and ensure more reliable performance despite unpredictable data availability.

\section{Lessions Learned}
% In this work, we found that the timestamp variation caused sampling irregularity (like sensor jitter) leaded to smaller HAR performance loss than random dropout (like sensor failure), this result indicates tha
The time-continuous neural network did not demonstrate obviously better performance against the sampling irregularity issues than the standard discrete-time ANNs. 
and the former also did not achieve as good an macro F1 score as the latter in the experiments on the regular dataset, 
Despite inputting additional timestamp information in the time-continuous neural network, which is not what we expected. This result emphasizes the necessity for developing new models or adapting existing ones to better accommodate irregular time-series data in this domain.
The discrete-time ANNs still obtain a robust performance against the sample irregularity caused by timestamp variation.
If the overall sampling rate is around 50 Hz, less than 1 \% performance is lost in the experiments on the MHEALTH and MOTIONSENSE datasets as \cref{tab:result_summary} shows, which is also out of our expectation.
The results reveal that the time-discrete ANNs also have good robustness against the sampling irregularity caused by the sensor jitter if the sensor data can be read at a relatively high sampling rate.
As the higher sampling rate may lead to hardware inefficiency, a trade-off between the model's robustness and efficiency should be considered during model design.

\section{Conclusion}
In this work, we explored the challenges of irregular sampling in HAR datasets through simulated scenarios using timestamp variations and random dropout. Comprehensive evaluations demonstrated that discrete-time neural networks exhibit robustness to minor sampling inconsistencies (sensor jitter). Additionally, despite their design, LTNNs (CFC\_Solver model) fail to effectively address irregular sampling, exposing limitations in their current formulation. These results underline a critical need for developing or adapting models specifically tailored to handle irregular time-series data, paving the way for more reliable and accurate HAR systems in real-world applications.

% \section*{Acknowledgment}

\bibliographystyle{IEEEtran}
\bibliography{IEEEfull}

% Generated by IEEEtran.bst, version: 1.14 (2015/08/26)
\begin{thebibliography}{10}
\providecommand{\url}[1]{#1}
\csname url@samestyle\endcsname
\providecommand{\newblock}{\relax}
\providecommand{\bibinfo}[2]{#2}
\providecommand{\BIBentrySTDinterwordspacing}{\spaceskip=0pt\relax}
\providecommand{\BIBentryALTinterwordstretchfactor}{4}
\providecommand{\BIBentryALTinterwordspacing}{\spaceskip=\fontdimen2\font plus
\BIBentryALTinterwordstretchfactor\fontdimen3\font minus \fontdimen4\font\relax}
\providecommand{\BIBforeignlanguage}[2]{{%
\expandafter\ifx\csname l@#1\endcsname\relax
\typeout{** WARNING: IEEEtran.bst: No hyphenation pattern has been}%
\typeout{** loaded for the language `#1'. Using the pattern for}%
\typeout{** the default language instead.}%
\else
\language=\csname l@#1\endcsname
\fi
#2}}
\providecommand{\BIBdecl}{\relax}
\BIBdecl

\bibitem{sipola2022artificial}
T.~Sipola, J.~Alatalo, T.~Kokkonen, and M.~Rantonen, ``Artificial intelligence in the iot era: A review of edge ai hardware and software,'' in \emph{2022 31st Conference of Open Innovations Association (FRUCT)}.\hskip 1em plus 0.5em minus 0.4em\relax IEEE, 2022, pp. 320--331.

\bibitem{singh2023survey}
R.~Singh, R.~Sukapuram, and S.~Chakraborty, ``A survey of mobility-aware multi-access edge computing: Challenges, use cases and future directions,'' \emph{Ad Hoc Networks}, vol. 140, p. 103044, 2023.

\bibitem{hasegawa2020smartphone}
T.~Hasegawa, ``Smartphone sensor-based human activity recognition robust to different sampling rates,'' \emph{IEEE Sensors Journal}, vol.~21, no.~5, pp. 6930--6941, 2020.

\bibitem{hamouda2020variable}
Y.~Hamouda and M.~Msallam, ``Variable sampling interval for energy-efficient heterogeneous precision agriculture using wireless sensor networks,'' \emph{Journal of King Saud University-Computer and Information Sciences}, vol.~32, no.~1, pp. 88--98, 2020.

\bibitem{coviello2020study}
G.~Coviello, G.~Avitabile, A.~Florio, and C.~Talarico, ``A study on imu sampling rate mismatch for a wireless synchronized platform,'' in \emph{2020 IEEE 63rd International Midwest Symposium on Circuits and Systems (MWSCAS)}.\hskip 1em plus 0.5em minus 0.4em\relax IEEE, 2020, pp. 229--232.

\bibitem{chinaev2021online}
A.~Chinaev, G.~Enzner, T.~Gburrek, and J.~Schmalenstroeer, ``Online estimation of sampling rate offsets in wireless acoustic sensor networks with packet loss,'' in \emph{2021 29th European Signal Processing Conference (EUSIPCO)}.\hskip 1em plus 0.5em minus 0.4em\relax IEEE, 2021, pp. 1110--1114.

\bibitem{vasconcelos2018data}
I.~L. Vasconcelos, I.~C. Martins, C.~M. Figueiredo, and A.~L. Aquino, ``A data sample algorithm applied to wireless sensor network with disruptive connections,'' \emph{Computer Networks}, vol. 146, pp. 1--11, 2018.

\bibitem{he2006sensor}
Y.~He and E.~K. Chong, ``Sensor scheduling for target tracking: A monte carlo sampling approach,'' \emph{Digital Signal Processing}, vol.~16, no.~5, pp. 533--545, 2006.

\bibitem{zhang2013distributed}
Y.~Zhang, S.~He, J.~Chen, Y.~Sun, and X.~S. Shen, ``Distributed sampling rate control for rechargeable sensor nodes with limited battery capacity,'' \emph{IEEE Transactions on Wireless Communications}, vol.~12, no.~6, pp. 3096--3106, 2013.

\bibitem{wang2010networked}
F.~Wang and J.~Liu, ``Networked wireless sensor data collection: issues, challenges, and approaches,'' \emph{IEEE Communications Surveys \& Tutorials}, vol.~13, no.~4, pp. 673--687, 2010.

\bibitem{ananthanarayanan2017real}
G.~Ananthanarayanan, P.~Bahl, P.~Bod{\'\i}k, K.~Chintalapudi, M.~Philipose, L.~Ravindranath, and S.~Sinha, ``Real-time video analytics: The killer app for edge computing,'' \emph{computer}, vol.~50, no.~10, pp. 58--67, 2017.

\bibitem{liu2024ikan}
M.~Liu, S.~Bian, B.~Zhou, and P.~Lukowicz, ``ikan: Global incremental learning with kan for human activity recognition across heterogeneous datasets,'' in \emph{Proceedings of the 2024 ACM International Symposium on Wearable Computers}, 2024, pp. 89--95.

\bibitem{bian2022state}
S.~Bian, M.~Liu, B.~Zhou, and P.~Lukowicz, ``The state-of-the-art sensing techniques in human activity recognition: A survey,'' \emph{Sensors}, vol.~22, no.~12, p. 4596, 2022.

\bibitem{bian2021capacitive}
S.~Bian and P.~Lukowicz, ``Capacitive sensing based on-board hand gesture recognition with tinyml,'' in \emph{Adjunct Proceedings of the 2021 ACM International Joint Conference on Pervasive and Ubiquitous Computing and Proceedings of the 2021 ACM International Symposium on Wearable Computers}, 2021, pp. 4--5.

\bibitem{phukan2022convolutional}
N.~Phukan, S.~Mohine, A.~Mondal, M.~S. Manikandan, and R.~B. Pachori, ``Convolutional neural network-based human activity recognition for edge fitness and context-aware health monitoring devices,'' \emph{IEEE Sensors Journal}, vol.~22, no.~22, pp. 21\,816--21\,826, 2022.

\bibitem{chen2021deep}
K.~Chen, D.~Zhang, L.~Yao, B.~Guo, Z.~Yu, and Y.~Liu, ``Deep learning for sensor-based human activity recognition: Overview, challenges, and opportunities,'' \emph{ACM Computing Surveys (CSUR)}, vol.~54, no.~4, pp. 1--40, 2021.

\bibitem{liu2018impact}
K.-C. Liu, C.-Y. Hsieh, S.~J.-P. Hsu, and C.-T. Chan, ``Impact of sampling rate on wearable-based fall detection systems based on machine learning models,'' \emph{IEEE Sensors Journal}, vol.~18, no.~23, pp. 9882--9890, 2018.

\bibitem{qi2013adasense}
X.~Qi, M.~Keally, G.~Zhou, Y.~Li, and Z.~Ren, ``Adasense: Adapting sampling rates for activity recognition in body sensor networks,'' in \emph{2013 IEEE 19th Real-Time and Embedded Technology and Applications Symposium (RTAS)}.\hskip 1em plus 0.5em minus 0.4em\relax IEEE, 2013, pp. 163--172.

\bibitem{rubanova2019latent}
Y.~Rubanova, R.~T. Chen, and D.~K. Duvenaud, ``Latent ordinary differential equations for irregularly-sampled time series,'' in \emph{Advances in Neural Information Processing Systems}, vol.~32, 2019.

\bibitem{hasani2021liquid}
R.~Hasani, M.~Lechner, A.~Amini, D.~Rus, and R.~Grosu, ``Liquid time-constant networks,'' in \emph{Proceedings of the AAAI Conference on Artificial Intelligence}, vol.~35, no.~9, 2021, pp. 7657--7666.

\bibitem{schafer2002missing}
J.~L. Schafer and J.~W. Graham, ``Missing data: our view of the state of the art,'' \emph{Psychological Methods}, vol.~7, no.~2, pp. 147--177, 2002.

\bibitem{nelson1997modeling}
J.~M. Nelson and D.~E. Myers, ``Modeling missing data in environmental monitoring using multiple imputation,'' \emph{Journal of Statistical Computation and Simulation}, vol.~58, no.~1, pp. 29--43, 1997.

\bibitem{geissler2024power}
D.~Gei{\ss}ler, B.~Zhou, M.~Liu, S.~Suh, and P.~Lukowicz, ``The power of training: How different neural network setups influence the energy demand,'' in \emph{International Conference on Architecture of Computing Systems}.\hskip 1em plus 0.5em minus 0.4em\relax Springer, 2024, pp. 33--47.

\bibitem{che2018recurrent}
Z.~Che, S.~Purushotham, K.~Cho, D.~Sontag, and Y.~Liu, ``Recurrent neural networks for multivariate time series with missing values,'' \emph{Scientific Reports}, vol.~8, no.~1, p. 6085, 2018.

\bibitem{chen2018neural}
R.~T.~Q. Chen, Y.~Rubanova, J.~Bettencourt, and D.~K. Duvenaud, ``Neural ordinary differential equations,'' in \emph{Advances in Neural Information Processing Systems}, vol.~31, 2018, pp. 6571--6583.

\bibitem{reiss2012pamap2}
A.~Reiss and D.~Stricker, ``Introducing a new benchmarked dataset for activity monitoring,'' in \emph{2012 16th international symposium on wearable computers}.\hskip 1em plus 0.5em minus 0.4em\relax IEEE, 2012, pp. 108--109.

\bibitem{banos2014mhealth}
O.~Banos, R.~Garcia, J.~A. Holgado-Terriza, M.~Damas, H.~Pomares, I.~Rojas, A.~Saez, and C.~Villalonga, ``mhealthdroid: a novel framework for agile development of mobile health applications,'' in \emph{Ambient Assisted Living and Daily Activities: 6th International Work-Conference, IWAAL 2014, Belfast, UK, December 2-5, 2014. Proceedings 6}.\hskip 1em plus 0.5em minus 0.4em\relax Springer, 2014, pp. 91--98.

\bibitem{malekzadeh2018motionsense}
M.~Malekzadeh, R.~G. Clegg, A.~Cavallaro, and H.~Haddadi, ``Protecting sensory data against sensitive inferences,'' in \emph{Proceedings of the 1st Workshop on Privacy by Design in Distributed Systems}, 2018, pp. 1--6.

\bibitem{ordonez2016deep}
F.~J. Ord{\'o}{\~n}ez and D.~Roggen, ``Deep convolutional and lstm recurrent neural networks for multimodal wearable activity recognition,'' \emph{Sensors}, vol.~16, no.~1, p. 115, 2016.

\bibitem{hasani2022closed}
R.~Hasani, M.~Lechner, A.~Amini, L.~Liebenwein, A.~Ray, M.~Tschaikowski, G.~Teschl, and D.~Rus, ``Closed-form continuous-time neural networks,'' \emph{Nature Machine Intelligence}, vol.~4, no.~11, pp. 992--1003, 2022.

\end{thebibliography}

\end{document}